\documentclass[twoside,twocolumn]{article}

\usepackage{blindtext} % Package to generate dummy text throughout this template 

\usepackage{graphicx}
\usepackage{subfigure}

\usepackage[sc]{mathpazo} % Use the Palatino font
\usepackage[T1]{fontenc} % Use 8-bit encoding that has 256 glyphs
\usepackage{microtype} % Slightly tweak font spacing for aesthetics

\usepackage[english]{babel} % Language hyphenation and typographical rules

\usepackage[hmarginratio=1:1,top=32mm,columnsep=20pt]{geometry} % Document margins
\usepackage[hang, small,labelfont=bf,up,textfont=it,up]{caption} % Custom captions under/above floats in tables or figures
\usepackage{booktabs} % Horizontal rules in tables

\usepackage{lettrine} % The lettrine is the first enlarged letter at the beginning of the text

\usepackage{enumitem} % Customized lists
\setlist[itemize]{noitemsep} % Make itemize lists more compact

\usepackage{abstract} % Allows abstract customization
\usepackage{titlesec} % Allows customization of titles
\renewcommand\thesection{\Roman{section}} % Roman numerals for the sections
\renewcommand\thesubsection{\roman{subsection}} % roman numerals for subsections
\titleformat{\section}[block]{\large\scshape\centering}{\thesection.}{1em}{} % Change the look of the section titles
\titleformat{\subsection}[block]{\large}{\thesubsection.}{1em}{} % Change the look of the section titles

\usepackage{fancyhdr} % Headers and footers
\usepackage{titling} % Customizing the title section

\usepackage{hyperref} % For hyperlinks in the PDF

\newcommand{\ra}{\rangle}
\newcommand{\la}{\langle}

\newcommand{\beq}{\begin{equation}}
\newcommand{\eeq}{\end{equation}}

\newcommand{\ball}{\begin{align}}
\newcommand{\eall}{\end{align}}

\newcommand{\beqar}{\begin{eqnarray}}
\newcommand{\eeqar}{\end{eqnarray}}

\newcommand{\ben}{\begin{enumerate}}
\newcommand{\een}{\end{enumerate}}

\pretitle{\begin{center}\huge\bfseries} % Article title formatting
\posttitle{\end{center}} % Article title closing formatting
\title{On the airborne aspect of COVID-19 coronovirus} % Article title
\author{%
\textsc{Navinder Singh, }\thanks{Theoretical Physics Division, Physical Research Laboratory, Ahmedabad, India. Cell Phone: +919662680605.} ~ and \textsc{Manpreet kaur}\thanks{Department of Pediatric and preventive dentistry, Ahmedabad Dental College, Ranchodpura, Ahmedabad. India. Cell Phone: +919409290203}  \\[1ex] % Your name
\normalsize \href{mailto:navinder.phy@gmail.com}{navinder.phy@gmail.com; drmanpreetkaur30@gmail.com} % Your email address
}
\date{\today} % Leave empty to omit a date
\renewcommand{\maketitlehookd}{%
\begin{abstract}
It is a widely accepted view that COVID 19 is either transmitted via surface contamination or via close contact of an un-infected person with an infected person. Surface contamination usually happens when infected water droplets from exhalation/sneeze/cough of COVID sick person settle on nearby surfaces. To curb this, social distancing and good hand hygiene advise is advocated by World health Organization (WHO). We argue that COVID 19 coronovirus can also be airborne in a puff cloud loaded with infected droplets generated by COVID sick person. An elementary calculation shows that a $5~\mu m$ respiratory infected droplet can remain suspended for about 9.0 minutes and a $2~\mu m$ droplet can remain suspended for about an hour! And social distancing advise of 3 feet by WHO and 6 feet by CDC (Centers for Disease Control and Prevention) may not be sufficient in some circumstances as discussed in the text.   
\end{abstract}
}

\begin{document}

% Print the title
\maketitle

%----------------------------------------------------------------------------------------
%	ARTICLE CONTENTS
%----------------------------------------------------------------------------------------

%%%%%%%%%%%%%%%%%%%%%%%%%%%%%%%%%%%%%%%%%%%%%%%%%%%%%%%%%%%%%%%%%%%%%%%

COVID-19 is a pandemic. It has infected over 2.2 million people worldwide, and over 1.5 lakh people have died due to it, as of 20th April 2020.  Most of the information related to COVID-19 can be found at WHO website\cite{who}, and at CDC website\cite{cdc}. In the next section we present a calculation with the aim to calculate timescale over which the virus in infected water droplets exhaled by a COVID sick person stays in air.

\section{A Newtonian-Stokes calculation (Neglecting Brownian motion)} 

Let $h$ be the typical height of a human being ($\sim 1.6~meter$). Human exhalation generates mucosalivary droplets of varied size from $0.5~\mu m$ to $12~\mu m$\cite{yang}. Consider the trajectory of a droplet. Time taken to fall from height $h$ using Newtonian mechanics {\it neglecting air drag} is given by $\sqrt{(2 h/g)}$, where $g$ is acceleration due to gravity. For $h=1.6 ~meters$ this time is roughly $0.6~sec$ (less than a second). In this case we say that droplets are not airborne and settle to nearby surfaces in a time less than a second.

But this picture is radically modified when air drag is taken into account. The equation of motion including drag force ($f_d$) for the vertical downward direction is given by

\beq
m\frac{dv(t)}{dt} = mg-f_d.
\label{0}
\eeq
Here $f_d$ Stokes' drag force given by $6\pi\eta r v(t)$. $\eta$ is the viscosity of air, $r$ is the radius of water droplet, and $v(t)$ is its downward velocity at time $t$. For the time being consider that initial ($t=0$) downward velocity is zero. The solution of the above equation immediately gives 
\begin{figure}[!h]
\begin{center}
\includegraphics[height=4cm]{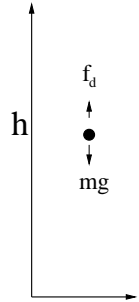}
\caption{Various forces acting on the water droplet.}
\end{center}
\end{figure}
\beq
v(t) = g\tau (1-e^{-t/\tau}),
\label{1}
\eeq
where $\tau = \frac{m}{6\pi\eta r}$. And the instantaneous height (measured from the top end) $v(t)=\frac{dh(t)}{dt}$ is given by
\beq
h(t) = g\tau (t + \tau(e^{-t/\tau}-1)).
\label{2}
\eeq
Let time taken to fall down be $t_0$ and $h(t_0) = h$ the typical height of a person.

Let us take radius of a typical water droplet to be $5~micro-meters$. Its mass will roughly be $5.25 \times 10^{-13}~kg$. Let us say ambient temperature is 25 degree centigrade. Air viscosity at this temperature is $\eta = 1.85 \times 10^{5}~Kg/(m-sec)$. Plugging these numbers we get $\tau = \frac{m}{6\pi\eta r} \simeq 3.0 \times 10^{-4}$ or $0.3~milli-secs$. For the shortest (drag free) time scale (0.6 seconds) the exponentials $e^{-t/\tau}$ in the expressions of $v(t)$ and $h(t)$ are extremely small. Thus $v(t)=v_{terminal}\simeq g\tau\simeq 3.0~mm/sec$ (from equation (\ref{1})) which is very low terminal speed! And that the terminal speed is reached even within millisecond time scale (as $\tau$ is in sub-milli-second timescale). 

The total time taken $t_0$ to fall down from height $h$ is $t_0 \simeq\frac{h}{g\tau}\simeq 544.0$ seconds, which is about 9 minutes! This is very surprising. In nil wind conditions, typical droplets of size $5~\mu m$ can take 9 minutes to settle down! Smaller droplets can take even more time! This time scale is inversely proportional to the square of the radius of droplet:
\beq
t_0 = 4.5 \left(\frac{\eta h}{g \rho r^2}\right),
\label{3}
\eeq
where $\rho$ is the density of water droplet. {\it $2~\mu m$ droplet can remain suspended for about an hour!}

If we consider initial downward component of velocity (as exhalation imparts some initial speed to droplets\footnote{Nasal breathing typically leads initial speeds of about $1.4 ~m/sec$.}) then the solution of the equation of motion leads to 

\beq
v(t) = v(0) e^{-t/\tau} + g\tau (1-e^{-t/\tau}).
\eeq 

Whatever be the initial speed it is going to decay in millisecond timescales as $\tau$ is very small in sub milli-second time scales. Also any transverse component of the water droplet will be dissipated away in such small time scales. It does not form a parabolic projectile trajectory as commonly visualized.

\section{Important physical situation related to social distancing}

As an implication of the above calculation, consider a situation like this: Consider that people are standing in a queue in front of a milk parlor or in front of an ATM for cash withdrawal waiting for their turn. Also consider that all are following social distancing norm and they stand 6 feet away from each other. Let us suppose that a person is infected but he/she is asymptomatic.\footnote{It has been seen that a person can be infectious even if she/he has no symptoms.} The person is standing at his/her place say for more than five minutes. Also consider a nil wind condition. This person will create an invisible cloud loaded with respiratory droplets of varied size. These droplets are all infected! After five minutes this person moves ahead to take his/her turn, and a person behind him/her takes his/her position. It will take couple of seconds to move 6 feet ahead. But the cloud loaded with infectious droplets is going to stay there! (a two micrometer droplet will remain suspended for about an hour, and smaller droplets take more time to settle). So this well person will enter this infectious cloud and can get infection, even if social distancing norms are obeyed!

Now, if the wind is flowing transverse to the people's queue, it will take away that infectious cloud, and it will disperse in the ambient air. At sufficiently large distances concentration of infectious droplet nuclei will be extremely low, and it will be no more dangerous. But if wind is flowing along the queue, then wind can transfer those infectious droplets to unprotected well persons.

\section{Life-time of the water droplets}

In the above discussion it is assumed that water droplets preserve their size. It is not true. They evaporate in short time depending upon the size of the droplet. However, a recent research by Lydia Bourouiba and collaborators show that life-times are radically modified as the droplets are in a very special environment of respiratory puff cloud which contains lots of water vapors\cite{lydia}. A thorough investigation of droplet life times and trajectories taking into account in this special environment is much needed (in equation (1) mass will then be a time depended parameter).  But, when an infected droplet evaporates,  it ends up in an infected droplet nuclei or an infected aerosol. These are sub-micron sized and take even more time to settle. Although the above mechanical-hydrodynamical approach cannot be applied, as aerosols dynamics will be stochastic and Brownian motion will be important in its transfer from one place to another, but it will take much more time to settle to ground! In addition, a turbulent wind can take it away. This points towards airborne nature of the disease. A discussion about the Brownian motion is given in the next section.

%%%%%%%%%%%%%%%%%%%%%%%%%%%%%%%%%%%%%%%%%%%%%%%%%%%%%%%%%%%%%%%%%
\section{Consideration of the Brownian motion}

In equation (1) we considered the systematic component (drag component $f_d$) of the force due to bombardment of droplet by air molecules and we neglected a random component of this force which is also due to bombardment of droplet by air molecules\footnote{Just imagine slow motion movie in which the droplet is being bombarded from all sides by air molecules. At one instant of time more molecules will bombard from left and the droplet will jerk towards left, and at the other instant of time this imbalance topples to some other direction. This leads to a random force.}. Thus the total force has two components: $F(t) = f_d + \xi(t)$. $\xi(t)$ is the random force whose average is zero $\la\xi(t)\ra =0$ but correlated in time. Simplest form of correlation is delta correlation $\la \xi(t)\xi(t^\prime)\ra = \Gamma\delta(t-t^\prime)$\cite{nav}.

This random force can be neglected for bigger droplets. The criterion is based on the value of Reynolds number. The fluctuating force $\xi(t)$ and the drag force $f_d$ are intimately related to each other through fluctuation dissipation theorem (as both originate from the common origin via molecular bombardment)\cite{nav}. Thus we need to compare drag force with another important force in the problem that is called the inertial force. The inertial force is the force required to change droplets momentum over a distance of its size $f_{inertial} \sim m v (v/r)$ (average rate of change of droplet's momentum over its size). The other force is the viscous force $f_{drag} = \eta r v$ (we neglect prefactors, as we are interested in order of magnitudes). Their ratio is called Reynolds number:

\beq
Re = \frac{f_{inertial}}{f_{drag}} = \frac{\rho r v}{\eta}.
\eeq

For $5~\mu m$ droplet drifting down with speed $3~mm/sec$, $Re \sim 0.8$. This is a borderline case. However, for Coronovirus (size $\sim 125 ~nm$) moving at this drift speed the Reynolds number turns out to be of the order of $10^{-2}$. This is completely a fluctuation dissipation dominated regime, and the virus will perform a Brownian motion.

Thus, the timescales obtained above by Newton-Stokes analysis are the minimum possible timescales (these are the lower bounds). That is, the $5~\mu m$ droplet will take more than 9 minutes to settle. Detailed investigation using Langevin or more sophisticated model is beyond the scope of this investigation.

\section{Advise to Public}
In view of the airborne nature of the disease to some extent it is advised that when someone is out of one's house to get essential goods at a frequently visited place (like a grocery store) one should not only wear a good mask for oral and nasal protection, but a protective gear for eyes (eye guard or eye shield) is also highly recommended to cover-up the airborne aspect of the disease discussed in this work.

\section{Summary}
We show within Newtonian-Stokes elementary mechanics that respiratory droplets does not form a quick parabolic projectile trajectories as commonly visualized. Stokes drag dominates, and they fall down much slowly. A five micron droplet takes about nine minutes to fall down from a height of 1.6 meters, and a droplet of two micron size takes about an hour to settle down! This clearly shows airborne nature of the disease. Even when they evaporate they end up forming a sub-micron sized infected aerosols which can stay airborne over much longer timescales. In view of the above factors, much more precautionary steps needs to be taken to stop spread of this dangerous virus.

%----------------------------------------------------------------------------------------
%	REFERENCE LIST
%----------------------------------------------------------------------------------------

%----------------------------------------------------------------------------------------

\end{document}